\documentclass[preprint,12pt]{elsarticle}

\usepackage{amsmath}
\usepackage{amssymb}

\usepackage{rotating}      
\usepackage{booktabs}      
\usepackage{multirow}
\usepackage{tabularx}
\usepackage[table]{xcolor}
\usepackage{ragged2e}
\usepackage{caption}
\usepackage{graphicx}
\usepackage[hyphens]{url}

\definecolor{tableheader}{HTML}{E0F2F7}
\definecolor{rowhighlight}{HTML}{F8F8F8}

\begin{document}

\begin{frontmatter}

  \title{Multi-Channel Feature Fusion and Monte Carlo Dropout for
  Uncertainty-Aware Diabetic Retinopathy Grading}

  \author[aff1]{Saksham Kumar\corref{cor1}}

  \cortext[cor1]{Corresponding author. Email: sakshamkumarprasad03@gmail.com (Saksham Kumar)}

  \affiliation[aff1]{%
    organization = {Department of Computer Science \& Engineering,\\
     Indian Institute of Technology Patna },
    city         = {Patna},
    state        = {Bihar},
    postcode     = {801106},
    country      = {India}}

  \begin{abstract}
    Automated five-stage diabetic retinopathy (DR) grading requires more
    than high accuracy alone. Medical-grade deployment calls for
    lesion-aware preprocessing, ordinal predictions, calibrated
    uncertainty, and explainability to support reliable diagnostic
    systems. We present a unified pipeline that addresses these
    requirements using a Ben~Graham--green-channel CLAHE feature
    representation, an EfficientNetV2-L ordinal regressor, and Monte
    Carlo dropout for uncertainty-driven referral. Grad-CAM provides
    visual explanations aligned with clinically relevant lesions.

    The proposed method achieves a QWK of 91.31\% on the APTOS-2019
    official test split, placing it within the near-perfect agreement
    band ($>80\%$). At a 20\% referral rate, 293 of 366 images are
    automatically graded with a QWK of 90.40\%. More complex cases are
    referred for specialist assessment, demonstrating a practical
    trade-off among grading quality, automation, and patient safety in
    robust, reliable, and deployment-ready medical diagnostic systems.
  \end{abstract}

  \begin{highlights}
  \item The pipeline combines grayscale Ben~Graham normalisation,
    green-channel CLAHE, an audited morphological proxy, and an
    EfficientNetV2-L ordinal regressor.
  \item The current quantitative results are explicitly treated as
    exploratory because threshold fitting and referral selection reused
    labels from the 366-image analysis subset.
  \item MC-dropout score variance is evaluated as a selective-prediction
    signal, while Grad-CAM is limited to qualitative visualisation rather
    than claimed lesion-level validation.
  \end{highlights}


\end{frontmatter}

\section{Introduction}
\label{sec:intro}

\begin{figure}[!htbp]
  \centering
  \includegraphics[width=\linewidth]{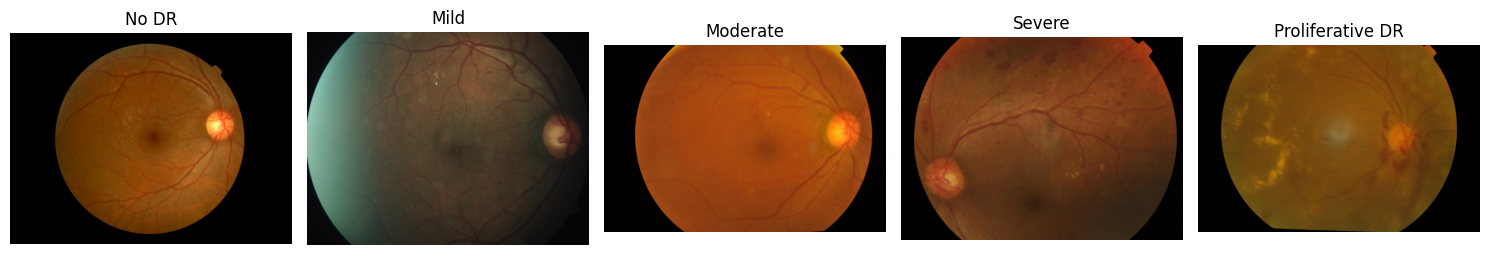}
  \caption{Representative fundus images for each of the five ICDRSS
  diabetic retinopathy severity stages (No~DR to Proliferative~DR).}
  \label{fig:all_stages}
\end{figure}

Diabetic retinopathy (DR) is a leading microvascular complication of
diabetes that progressively damages retinal capillaries and causes
irreversible vision loss if left untreated~\cite{Leasher2016Global2010}.
Chronic hyperglycaemia weakens vessel walls, producing microaneurysms,
haemorrhages, macular oedema, and pathological neovascularisation.
The disease advances through severity stages from mild
non-proliferative DR (Mild-NPDR) to proliferative DR (PDR), all of
which can be assessed from colour fundus photography
(Fig.~\ref{fig:all_stages}).

Population-level screening is essential, yet manual grading is
resource-intensive and subject to inter-observer variability among
clinicians~\cite{DiabeticScreening}. Digital fundus photography and
tele-ophthalmology have expanded coverage~\cite{Welch2024DiabeticOphthalmology},
but scaling these programmes requires reliable automated grading with
clinically meaningful signals.

Deep learning (DL)-based medical image analysis has demonstrated
strong performance in multi-stage DR grading~\cite{Chikumba2023DeepReview}.
However, a high point estimate alone is insufficient to establish
clinical utility. A candidate DR grading system should therefore be
evaluated for the following properties:

\begin{enumerate}
  \item \textbf{Robust preprocessing} that handles illumination,
    contrast, and device variability whilst retaining
    pathological cues.
  \item \textbf{Ordinal-aware predictions} that respect the progressive
    severity structure of the ICDRSS classification scale.
  \item \textbf{Uncertainty and selective-prediction performance}
    assessed with calibration, error-detection, and risk--coverage
    metrics.
  \item \textbf{Explanation validity} assessed for localisation and
    faithfulness rather than by visual impression alone.
\end{enumerate}

Prior studies have already examined uncertainty-informed referral for
DR detection and uncertainty-aware five-class grading
\cite{Leibig2017,Jaskari2022}. The present work studies a specific
combination of preprocessing, ordinal regression, MC dropout, and
Grad-CAM; it does not claim to be the first uncertainty-aware DR
pipeline.

\subsection{Dataset}
\label{subsec:dataset}

The APTOS-2019 dataset~\cite{Kaggle2019APTOSKaggle}, provided by
the Asia Pacific Tele-Ophthalmology Society and collected by Aravind
Eye Care (India), comprises 3{,}662 labelled fundus images captured
under diverse acquisition conditions. Labels follow the International
Clinical Diabetic Retinopathy Severity Scale
(ICDRSS)~\cite{2012InternationalOphthalmology}: No DR (0), Mild NPDR
(1), Moderate NPDR (2), Severe NPDR (3), and Proliferative DR (4).
Following the labelled derivative distributed by Herrero
\cite{Herrero2019}, the 3{,}662-image labelled pool is divided into
2{,}930 training, 366 validation, and 366 nominal test images. The last
subset is not the official Kaggle competition test set; it is a
held-out subset of the public labelled pool. Moreover, because its labels
were reused during threshold and referral selection, it is called the
\emph{analysis subset} below rather than a held-out test set. Its grade
counts are 199 No-DR, 30 mild-NPDR, 87 moderate-NPDR, 17 severe-NPDR,
and 33 PDR images. The available experiment record does not document
the split-generation seed, stratification procedure, or duplicate- and
patient-level leakage checks; these must be recorded for a confirmatory
rerun.

\subsection{Contributions}
\label{subsec:contributions}

This work investigates the following components:

\begin{enumerate}
  \item A \textbf{three-channel preprocessing} pipeline combining
    grayscale Ben~Graham illumination normalisation, green-channel
    CLAHE, and a morphological proxy map.
  \item An \textbf{ordinal regression framework} using EfficientNetV2-L
    with Huber loss and validation-fitted ordered grade thresholds.
  \item A \textbf{selective-prediction analysis} based on MC-dropout
    score variance, with explicit reporting of accepted high-grade
    failure cases.
  \item \textbf{Qualitative Grad-CAM visualisations} of positive
    influence on the scalar regression output.
  \item An \textbf{evaluation audit} that identifies the current leakage
    and specifies a locked validation/test protocol for future reporting.
\end{enumerate}

\section{Related Work}
\label{sec:relwork}

\subsection{Transfer Learning and CNN Architectures}

Transfer learning with pre-trained convolutional neural networks
(CNNs) is the dominant paradigm for DR classification.
Dixit et al.~\cite{Dixit2025FundusBlock} proposed EfficientNet-B3
augmented with squeeze-and-excitation (SE) blocks, achieving 88.44\%
accuracy and 85.39\% QWK on APTOS-2019; the channel-wise attention
from SE blocks amplifies DR-relevant features.
Singh et al.~\cite{Singh2024DiabeticModel} demonstrated that
extensive preprocessing significantly improves EfficientNet-B0
classification by sharpening DR-relevant structures.
Shakibania et al.~\cite{Shakibania2024DualRetinopathy} proposed a
dual-branch architecture combining ResNet-50 and EfficientNet-B0 as
parallel feature extractors, achieving 89.6\% accuracy in five-stage
grading (91.90\% QWK) on APTOS-2019 by merging the complementary
strengths of both backbones.

\subsection{Ensemble and Hierarchical Methods}

Jian et al.~\cite{Jian2023Triple-DRNet:Images} introduced
Triple-DRNet, a triple-cascade ResNet that decomposes five-class
grading hierarchically, achieving 92.08\% accuracy and 93.62\% QWK
on their APTOS-2019 evaluation protocol.
Rafid et al.~\cite{Rafid2024AnFramework} combined multiple deep CNNs
in an ensemble framework for early-stage DR diagnosis (82.74\%
accuracy), while Bodapati et al.~\cite{Bodapati2024Self-adaptivePrediction}
reported 88.58\% QWK with a self-adaptive meta-learner incorporating
dual-attention and spatial correlation modules.

\subsection{Attention Mechanisms}

Farag et al.~\cite{Farag2022CBAM} integrated DenseNet-169 with the
Convolutional Block Attention Module (CBAM), scoring 82\% five-class
accuracy and 88.80\% QWK on APTOS-2019; CBAM refocuses the model on
discriminative spatial and channel features.
A multi-head self-attention CNN (MHSA-CNN)~\cite{Abraham2026} reaches
91.85\% QWK on the same benchmark.
Lalithadevi et al.~\cite{Lalithadevi2024DiabeticTechnique} proposed
OptiDex (NASNet-Mobile + enhanced Cat Swarm optimisation + XAI),
achieving 97.65\% accuracy with integrated explainability---an
important step towards transparent diagnostic decision-making.

\subsection{Uncertainty-Aware DR Classification and Referral}

Leibig et al.~\cite{Leibig2017} evaluated dropout-based uncertainty
for DR detection and used it to rank cases for decision referral.
Jaskari et al.~\cite{Jaskari2022} subsequently studied approximate
Bayesian neural networks for both binary and five-class DR
classification, including clinical data and reject-option analysis.
These studies establish that uncertainty-informed DR referral and
five-class uncertainty estimation predate the present pipeline. They
also motivate evaluation with error-detection, calibration, and
risk--coverage measures rather than interpreting variance magnitude
alone as calibrated uncertainty.

\subsection{Contrastive and Non-Standard Approaches}

Islam et al.~\cite{Islam2022ApplyingImages} applied supervised
contrastive learning (SCL) with an Xception encoder and CLAHE
enhancement and reported
84.36\% five-class accuracy.
Mohsen et al.~\cite{Mohsen2025} introduced RadFuse, which fuses
non-linear RadEx (Radon-based) sinogram representations with fundus
images to achieve 93.24\% QWK and 87.07\% accuracy for five-stage
grading on APTOS-2019.
Oulhadj et al.~\cite{Oulhadj2022DiabeticRegistration} combined
deformable image registration with a multi-CNN voting ensemble
(75\% QWK), while Sikder et al.~\cite{Sikder2021SeverityImages}
used GLCM and histogram features with XGBoost for competitive
classical machine-learning performance.

\subsection{QWK Benchmarks on APTOS-2019}

Table~\ref{tab:qwk_litreview} summarises reported QWK scores for
five-stage DR grading on
APTOS-2019. Because split construction, tuning data, and evaluation
protocols differ among papers, the values are retained only as
descriptive literature context and are not ranked against the current
exploratory analysis.

\begin{table}[!htbp]
  \centering
  \caption{Selected QWK values reported for five-stage DR grading on
  APTOS-2019. Values are not directly comparable unless the dataset
  split and model-selection protocol match.}
  \label{tab:qwk_litreview}
  \begin{tabular}{llc}
    \toprule
    \textbf{Method} & \textbf{Year} & \textbf{QWK (\%)} \\
    \midrule
    DenseNet-169 + CBAM~\cite{Farag2022CBAM}           & 2022 & 88.80 \\
    ResNet-50 + EfficientNet-B0~\cite{Shakibania2024DualRetinopathy} & 2024 & 91.90 \\
    MHSA-CNN~\cite{Abraham2026}                        & 2026 & 91.85 \\
    RadFuse~\cite{Mohsen2025}                          & 2025 & 93.24 \\
    Triple-DRNet~\cite{Jian2023Triple-DRNet:Images}    & 2023 & \textbf{93.62} \\
    \bottomrule
  \end{tabular}
\end{table}

\subsection{Identified Gaps}
\label{subsec:gaps}

The literature indicates several evaluation needs that motivate the
present analysis:

\begin{enumerate}
  \item Dataset splits and threshold-selection protocols must be stated
    before raw QWK values can be compared.
  \item Referral rules require a validation-fixed threshold and
    accepted-case failure analysis, particularly for severe NPDR and PDR.
  \item Morphological feature channels require ablation and lesion-level
    validation before they can be interpreted as lesion detectors.
  \item Ordinal thresholds must be constrained, fitted without test-label
    access, and stored with the selected checkpoint.
  \item Attribution maps require localisation and faithfulness testing
    before supporting clinical or regulatory claims.
\end{enumerate}

The comprehensive survey of representative methods is reproduced in
Tables~\ref{tab:survey_part1} and~\ref{tab:survey_part2}.

\begin{table}[!htbp]
  \centering
  \captionof{table}{Survey of recent works in diabetic retinopathy
  detection -- Part~1.}
  \label{tab:survey_part1}
  \begin{sideways}
    \begin{tabularx}{0.95\textheight}{@{}
        >{\bfseries}p{3cm}
        >{\RaggedRight}X
        >{\RaggedRight}X
        >{\RaggedRight}X
      >{\Centering}p{4cm}@{}}
      \toprule
      \rowcolor{tableheader}
      \textbf{Paper} & \textbf{Methodology} &
      \textbf{Preprocessing} & \textbf{Classification Type} &
      \textbf{Evaluation Metric} \\
      \midrule
      \cite{Dixit2025FundusBlock} &
      EfficientNet-B3 + SE blocks &
      Image enhancement, augmentation &
      5-class classification &
      88.44\% Acc, 85.39\% QWK \\
      \rowcolor{rowhighlight}
      \cite{Lalithadevi2024DiabeticTechnique} &
      OptiDex (NASNet-Mobile + enhanced Cat Swarm + XAI) &
      Morphological processing, NAS optimisation &
      Detection and severity classification &
      97.65\% Acc \\
      \cite{Singh2024DiabeticModel} &
      EfficientNet-B0 with preprocessing strategies &
      Extensive preprocessing &
      DR detection &
      83\% Acc \\
      \rowcolor{rowhighlight}
      \cite{Bodapati2024Self-adaptivePrediction} &
      Self-stacking ensemble &
      Lesion isolation with attention &
      5-class classification &
      88.58\% QWK \\
      \cite{Shakibania2024DualRetinopathy} &
      Dual-branch transfer learning (ResNet-50 + EfficientNet-B0) &
      Augmentation, quality enhancement &
      Binary + 5-class grading &
      89.6\% Acc, 91.90\% QWK \\
      \rowcolor{rowhighlight}
      \cite{Rafid2024AnFramework} &
      Ensemble of deep CNNs &
      Standard preprocessing &
      Multi-class classification &
      82.74\% Acc \\
      \cite{Jian2023Triple-DRNet:Images} &
      Triple-DRNet: triple-cascade ResNet &
      Standard preprocessing &
      5-class cascade classification &
      92.08\% Acc, 93.62\% QWK \\
      \bottomrule
    \end{tabularx}
  \end{sideways}
\end{table}

\clearpage

\begin{table}[!htbp]
  \centering
  \captionof{table}{Survey of recent works in diabetic retinopathy
  detection -- Part~2.}
  \label{tab:survey_part2}
  \begin{sideways}
    \begin{tabularx}{0.95\textheight}{@{}
        >{\bfseries}p{3cm}
        >{\RaggedRight}X
        >{\RaggedRight}X
        >{\RaggedRight}X
      >{\Centering}p{4cm}@{}}
      \toprule
      \rowcolor{tableheader}
      \textbf{Paper} & \textbf{Methodology} &
      \textbf{Preprocessing} & \textbf{Classification Type} &
      \textbf{Evaluation Metric} \\
      \midrule
      \cite{Raiaan2023AImages} &
      10-layer shallow ResNet &
      Lightweight preprocessing &
      Multi-class classification &
      98.65\% Acc \\
      \rowcolor{rowhighlight}
      \cite{Kasim2023EnsembleRetinopathy} &
      Ensemble with optimised transfer learning &
      Feature extraction and selection &
      Early-stage diagnosis &
      88.95\% Acc \\
      \cite{Islam2022ApplyingImages} &
      Supervised contrastive learning + Xception &
      CLAHE enhancement &
      Binary + 5-class &
      98.36\% Acc (binary), 84.36\% Acc (5-class) \\
      \rowcolor{rowhighlight}
      \cite{Farag2022CBAM} &
      DenseNet-169 + CBAM attention &
      Standard preprocessing &
      Binary + 5-class severity grading &
      97\% Acc (binary), 82\% Acc (5-class), 88.80\% QWK \\
      \cite{Mohsen2025} &
      RadFuse: non-linear RadEx (Radon) + fundus fusion &
      RadEx sinogram + fundus preprocessing &
      Binary + 5-class grading &
      99.09\% Acc (binary), 87.07\% Acc, 93.24\% QWK (5-class) \\
      \rowcolor{rowhighlight}
      \cite{Oulhadj2022DiabeticRegistration} &
      Deformable registration + multi-CNN voting ensemble &
      Deformable registration &
      5-class grading &
      85.28\% Acc, 75\% QWK \\
      \cite{Sikder2021SeverityImages} &
      GLCM + histogram features + XGBoost ensemble &
      Feature extraction (GLCM, histogram) &
      5-class grading &
      94\% Acc \\
      \bottomrule
    \end{tabularx}
  \end{sideways}
\end{table}

\section{Methodology}
\label{sec:method}

The pipeline is organised around preprocessing, ordinal regression,
MC-dropout score sampling, selective referral, and qualitative
visualisation.

\begin{figure}[!htbp]
  \centering
  \includegraphics[width=\linewidth]{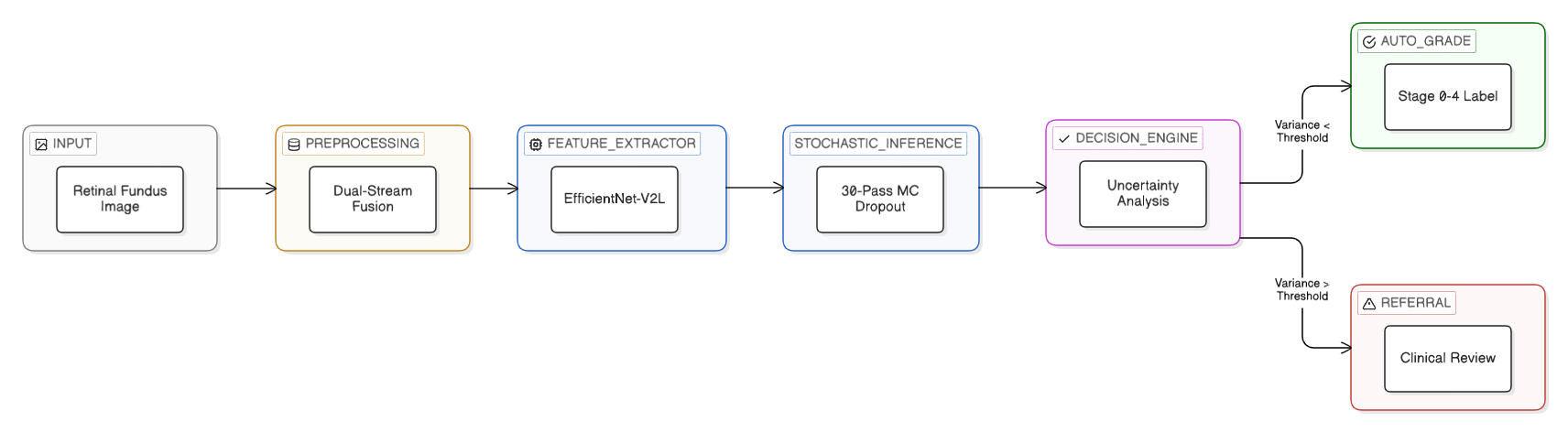}
  \caption{End-to-end analysis pipeline. Ordinal grade thresholds and
  the referral threshold must be fixed on validation data before locked
  evaluation.}
  \label{fig:system_pipeline}
\end{figure}

Fig.~\ref{fig:system_pipeline} illustrates the end-to-end architecture.

\subsection{Preprocessing: Audited Multi-Channel Representation}
\label{subsec:preproc}

\begin{figure}[!htbp]
  \centering
  \includegraphics[width=\linewidth]{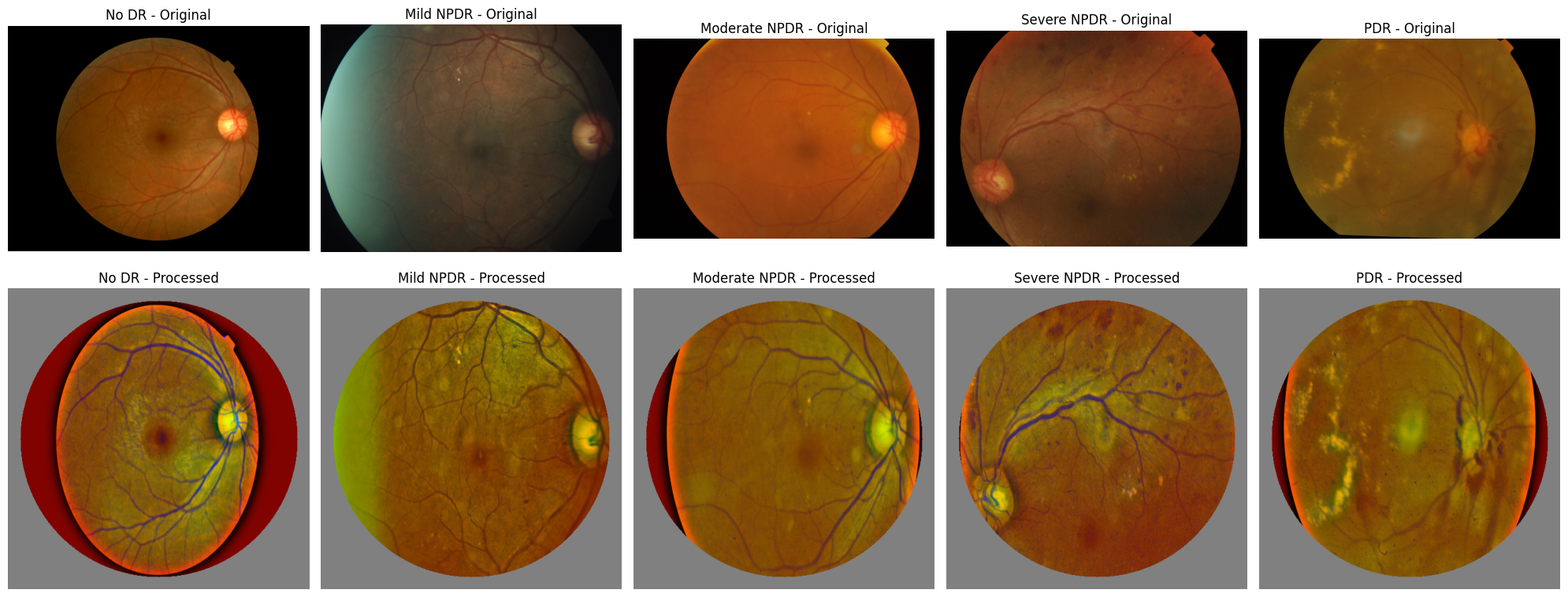}
  \caption{Feature-enhanced images produced by the three-channel
    preprocessing pipeline. From left: original fundus,
    grayscale Ben~Graham channel, Green-CLAHE channel, implemented
  dark-structure proxy, and composite input.}
  \label{fig:bg_processed}
\end{figure}

\begin{figure}[!htbp]
  \centering
  \includegraphics[width=\linewidth]{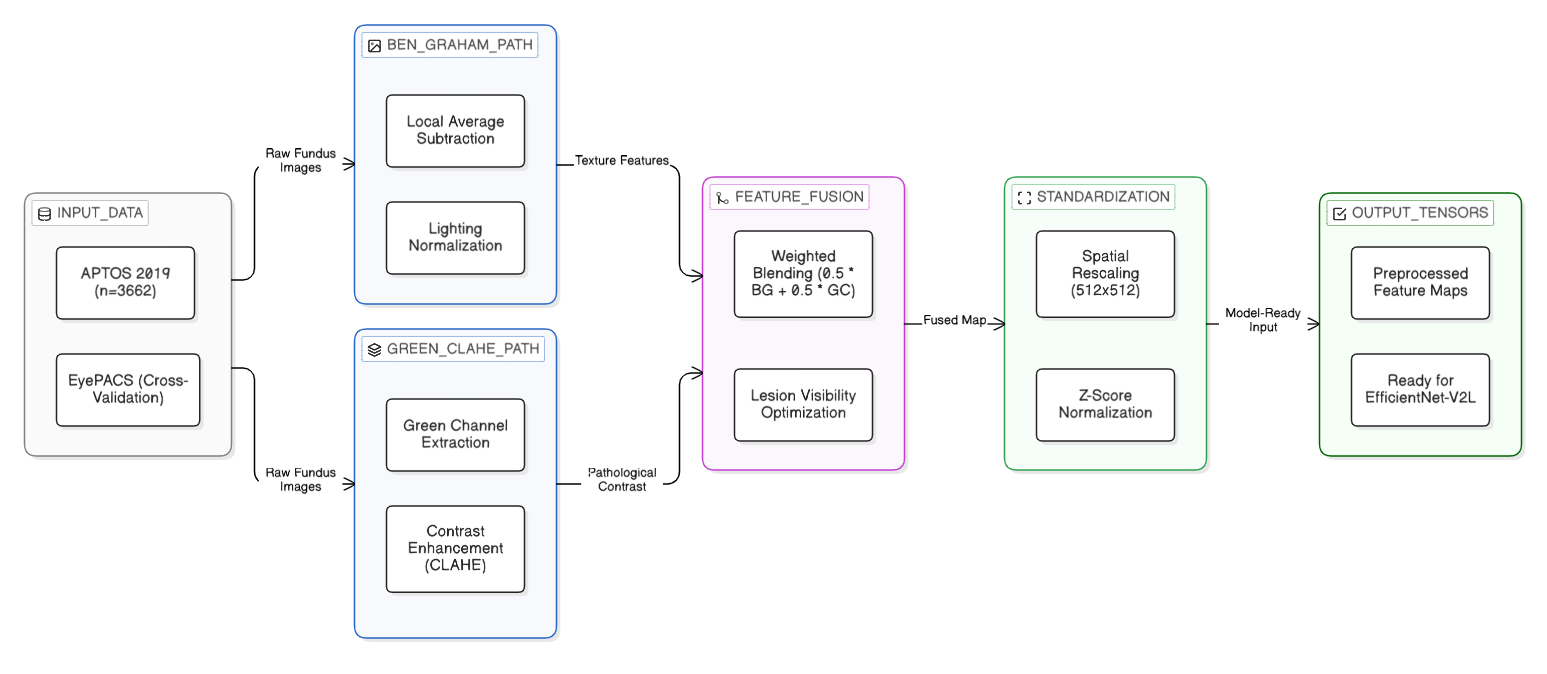}
  \caption{Feature-enhanced images produced by the three-channel
    preprocessing pipeline. From left: original fundus,
    grayscale Ben~Graham channel, Green-CLAHE channel, implemented
  dark-structure proxy, and composite input.}
  \label{fig:preprocessing_pipeline}
\end{figure}

Each RGB fundus image is resized to $512{\times}512$ and converted
into a three-channel feature-enhanced image (Fig.~\ref{fig:bg_processed};
pipeline illustrated in Fig.~\ref{fig:preprocessing_pipeline}):

\begin{itemize}
  \item \textbf{Channel~0 -- Ben~Graham Illumination Normalisation}
    ($\sigma = 10$): the Ben~Graham-normalised image is converted to
    grayscale, yielding the first scalar channel.
  \item \textbf{Channel~1 -- Green-Channel CLAHE}
    ($\mathrm{clipLimit} = 2.0$, $\mathrm{tileGridSize} = 8{\times}8$):
    enhances local contrast in the green channel, which carries
    the strongest retinal signal, amplifying microaneurysms and
    vessel detail.
  \item \textbf{Channel~2 -- Morphological Dark-Structure Proxy}:
    uses a black-hat response with a $15{\times}15$ elliptical
    structuring element, followed by Gaussian smoothing
    ($5{\times}5$). It is not treated as a validated lesion detector.
\end{itemize}

An audit of the supplied preprocessing code found that its two stated
morphological terms are redundant. For grayscale image $I$ and the same
symmetric structuring element $S$, it computes
\begin{equation}
  B(I)=(I\mathbin{\bullet}S)-I,
  \qquad
  T(255-I)=(255-I)-[(255-I)\mathbin{\circ}S]=B(I),
  \label{eq:morphology_audit}
\end{equation}
where $\bullet$ and $\circ$ denote closing and opening. The implemented
weighted sum, $0.6B(I)+0.6T(255-I)$, is therefore essentially a scaled
black-hat response and contains no separate bright-structure term. A
separate per-image normalisation convention is not defined in the
current manuscript record and must be documented in the rerun. A
corrected complementary channel would instead combine $B(I)$ with the
white top-hat $T(I)=I-(I\mathbin{\circ}S)$ on the original grayscale
image. That correction changes the model input and consequently requires
preprocessing, training, threshold selection, and evaluation to be
rerun. The present results are therefore not used to claim a
haemorrhage--exudate fusion benefit.

Pixels outside the retinal disc (radius scale = 0.46) are set to a
neutral value of 128, eliminating camera-edge artefacts. At training
time images are resized to $384{\times}384$ with spatial-only
augmentation (horizontal flip, vertical flip, random $90^\circ$ rotation)
and normalised with ImageNet statistics.

Preprocessing hyperparameters are summarised in
Table~\ref{tab:processing_pipeline}.

\begin{table}[!htbp]
  \centering
  \caption{Preprocessing pipeline parameters and rationale.}
  \label{tab:processing_pipeline}
  \begin{tabularx}{\linewidth}{>{\RaggedRight\arraybackslash}p{0.20\linewidth}>{\RaggedRight\arraybackslash}X>{\RaggedRight\arraybackslash}p{0.24\linewidth}}
    \toprule
    \textbf{Step} & \textbf{Rationale} & \textbf{Parameter value} \\
    \midrule
    Ben~Graham normalisation &
    Corrects global illumination to improve vessel visibility and
    cross-device consistency. &
    $\sigma = 10$ \\
    Green-channel CLAHE &
    Enhances local contrast for micro-lesions and vessels in the
    most informative channel. &
    \texttt{clipLimit}: 2.0; \texttt{tileGridSize}: $8{\times}8$ \\
    Dark-structure proxy &
    Produces the implemented black-hat-equivalent response; no separate
    bright-lesion term is present. &
    weights: $0.6+0.6$; kernel: $15{\times}15$; blur: $5{\times}5$ \\
    Disc mask + resize &
    Masks non-retinal background; standardises spatial dimensions. &
    Radius scale: 0.46; size: $512{\times}512$ \\
    \bottomrule
  \end{tabularx}
\end{table}

\subsection{Model Architecture: Ordinal Regression and Thresholding}
\label{subsec:model}

\begin{figure}[!htbp]
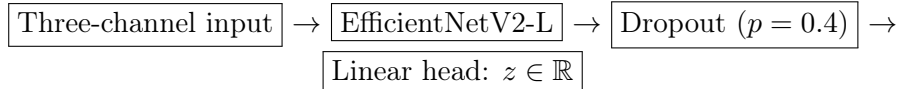

  \centering
  \small
  \fbox{Three-channel input}
  $\rightarrow$
  \fbox{EfficientNetV2-L}
  $\rightarrow$
  \fbox{Dropout ($p=0.4$)}
  $\rightarrow$
  \fbox{Linear head: $z\in\mathbb{R}$}
  \caption{EfficientNetV2-L ordinal regression architecture. The
  scalar output is unbounded before ordered thresholding.}
  \label{fig:model_architecture}
\end{figure}

The preprocessed three-channel image is fed into an
EfficientNetV2-L backbone pre-trained on ImageNet-21k and fine-tuned
on ImageNet-1k
(\texttt{tf\_efficientnetv2\_l.in21k\_ft\_in1k})~\cite{tan2021efficientnetv2},
with a dropout layer (rate = 0.4) and a single linear regression
head (Fig.~\ref{fig:model_architecture}). The head produces an
unbounded raw score $z=f_\phi(x)\in\mathbb{R}$; no bounded link or
clipping operation is applied before the loss or thresholding.

\paragraph{Loss function.}
Huber loss ($\delta = 1.0$) is adopted for its robustness to noisy
labels whilst remaining aligned with the ordinal evaluation objective
(QWK). Unlike cross-entropy, Huber loss penalises large severity
errors more heavily than small ones, consistent with the clinical
cost structure of DR grading.

\paragraph{Ordinal threshold optimisation.}
For a confirmatory evaluation, four ordered thresholds must be fitted
on validation predictions only. Ordering can be enforced by optimising
unconstrained parameters $a_k$ through
\begin{equation}
  \theta_1=a_1,\qquad
  \theta_k=\theta_{k-1}+\operatorname{softplus}(a_k),
  \quad k=2,3,4.
  \label{eq:ordered_thresholds}
\end{equation}
The selected checkpoint and thresholds are then frozen. A raw score
$z$ is assigned to grade $\hat g$ according to

\begin{equation}
  \hat{g} = \sum_{k=1}^{4} \mathbf{1}[z > \theta_k],
  \quad \theta_1 < \theta_2 < \theta_3 < \theta_4.
  \label{eq:thresholding}
\end{equation}

Thus a score exactly equal to a threshold remains in the lower grade.
The audited experiment did not enforce this parameterisation and reused
analysis-subset labels during fitting; its values are reported only as
post-hoc exploratory results in Section~\ref{subsec:grading_perf}.

\paragraph{Training configuration.}
Training hyperparameters are listed in Table~\ref{tab:model_params}.
Class imbalance is addressed by \texttt{WeightedRandomSampler}, which
up-samples minority grades during each epoch. Float16 automatic mixed
precision and gradient accumulation over four steps yield an effective
batch size of 128.

\begin{table}[!htbp]
  \centering
  \caption{Training hyperparameters for the EfficientNetV2-L
  ordinal regressor.}
  \label{tab:model_params}
  \begin{tabularx}{\linewidth}{>{\RaggedRight\arraybackslash}p{0.28\linewidth}>{\RaggedRight\arraybackslash}X}
    \toprule
    \textbf{Hyperparameter} & \textbf{Value} \\
    \midrule
    Backbone      & EfficientNetV2-L (\texttt{tf\_efficientnetv2\_l.in21k\_ft\_in1k}) \\
    Input size    & $384{\times}384$ (from $512{\times}512$ preprocessed features) \\
    Devices       & Two NVIDIA Tesla T4 GPUs with \texttt{DataParallel} \\
    Batch size    & 16 per GPU (32 global) \\
    Gradient accumulation & 4 steps (effective batch: 128) \\
    Epochs        & 20 \\
    Optimiser     & AdamW \\
    LR schedule   & Warmup (3 epochs) + cosine decay ($10^{-6}$ $\to$ $10^{-4}$ $\to$ $10^{-6}$) \\
    Loss function & Huber ($\delta = 1.0$) \\
    Dropout rate  & 0.4 \\
    Class balancing & \texttt{WeightedRandomSampler} \\
    Mixed precision & Enabled (float16 automatic mixed precision) \\
    \bottomrule
  \end{tabularx}
\end{table}

\paragraph{Computational infrastructure.}
The result-producing notebook records training on two NVIDIA Tesla T4
GPUs using PyTorch \texttt{DataParallel}. The exact code commit,
dataset version, checkpoint identifier, preprocessing host, and random
seeds are not preserved in the current manuscript artifacts and must be
archived with the confirmatory rerun.

\subsection{Selective Referral via Monte Carlo Dropout}
\label{subsec:mcdropout}

Monte Carlo (MC) dropout~\cite{Gal2016Dropout} retains the dropout
layer active at inference and performs $T = 30$ stochastic forward
passes per image. The per-sample predictive variance is:

\begin{equation}
  \sigma^2_{\mathrm{MC}} = \frac{1}{T}\sum_{t=1}^{T}z_t^2
  - \left(\frac{1}{T}\sum_{t=1}^{T}z_t\right)^2,
  \label{eq:mc_variance}
\end{equation}

where $z_t\in\mathbb{R}$ is the raw scalar regression output on
pass $t$.
Samples exceeding a variance threshold $\tau$ are flagged for
specialist review rather than auto-graded. For confirmatory evaluation,
$\tau$ must be chosen on validation data using a prespecified objective
and then applied numerically unchanged to the held-out set. The
zero-referral endpoint retains every case, equivalently
$\tau=+\infty$. The sampled scalar variance is called
\emph{MC-dropout score variance}; it is not assumed to be calibrated or
to include aleatoric uncertainty.

\subsection{Explainability via Grad-CAM}
\label{subsec:gradcam}

Gradient-weighted Class Activation Maps
(Grad-CAM)~\cite{Selvaraju2017GradCAM} are computed by
back-propagating gradients of the scalar regression output through
the last convolutional block of EfficientNetV2-L. The resulting
heatmaps are overlaid on the saved preprocessed three-channel feature
image. They show regions with positive influence on the scalar output,
not class-specific evidence for the thresholded grade. The maps are
used only as qualitative visualisations pending lesion-localisation,
perturbation, and reader-validation studies.

\section{Results and Discussion}
\label{sec:results}

\subsection{Evaluation Metric: Quadratic Weighted Kappa}
\label{subsec:qwk}

The primary evaluation metric is Quadratic Weighted Kappa
(QWK)~\cite{Cohen1968}, which gives larger weights to larger ordinal
disagreements. For five grades, let $i,j\in\{0,\ldots,K-1\}$ with
$K=5$. Let $O_{i,j}$ be the observed joint proportions, normalised so
that $\sum_{i,j}O_{i,j}=1$, and let $E_{i,j}$ be the outer product of
the corresponding true- and predicted-grade marginal proportions.

\begin{equation}
  \mathrm{QWK} = 1 - \frac{\sum_{i,j} W_{i,j}\,O_{i,j}}
  {\sum_{i,j} W_{i,j}\,E_{i,j}},
  \label{eq:qwk}
\end{equation}

\noindent where $W_{i,j}=(i-j)^2/(K-1)^2$. The verbal bands in
Table~\ref{tab:QWK_ScoreTable} are the generic Landis--Koch observer-
agreement heuristic~\cite{Landis1977}; they are not part of ICDRSS and
do not define a clinical deployment threshold.

\begin{table}[!htbp]
  \centering
  \caption{Landis--Koch heuristic labels for kappa values. These
  descriptive bands are not clinical acceptance criteria.}
  \label{tab:QWK_ScoreTable}
  \begin{tabular}{ll}
    \toprule
    \textbf{QWK score range} & \textbf{Interpretation} \\
    \midrule
    $< 0$       & Worse than chance / systematic disagreement \\
    $[0.00,0.20]$  & Slight agreement \\
    $(0.20,0.40]$  & Fair agreement \\
    $(0.40,0.60]$  & Moderate agreement \\
    $(0.60,0.80]$  & Substantial agreement \\
    $(0.80,1.00]$  & Almost-perfect agreement \\
    \bottomrule
  \end{tabular}
\end{table}

\subsection{Grading Performance}
\label{subsec:grading_perf}

An implementation audit showed that the current experiment does not
provide a held-out test estimate. The evaluation routine fitted
thresholds when called on the 366-image analysis subset, so the quoted
QWK of 0.9084 already used its labels. A subsequent grid search over
$\theta_3$ and $\theta_4$ on the same labels produced 0.9125, and the
same analysis-selected thresholds produced 0.9131 when applied to the
MC-dropout means. These values are dependent post-hoc analyses and are
not evidence of successive generalisation improvements.

\begin{table}[!htbp]
  \centering
  \caption{Exploratory per-grade results reconstructed from the current
  confusion matrix. Thresholds were selected using labels from the same
  analysis subset, so the values are not held-out estimates.}
  \label{tab:model_stats}
  \begin{tabular}{lrrr}
    \toprule
    \textbf{True grade} & \textbf{Support} & \textbf{Correct} & \textbf{Recall} \\
    \midrule
    No DR         & 199 & 194 & 0.975 \\
    Mild NPDR     & 30  & 18  & 0.600 \\
    Moderate NPDR & 87  & 50  & 0.575 \\
    Severe NPDR   & 17  & 12  & 0.706 \\
    PDR           & 33  & 12  & 0.364 \\
    \bottomrule
  \end{tabular}
\end{table}

\begin{figure}[!htbp]
  \centering
  \includegraphics[width=\linewidth]{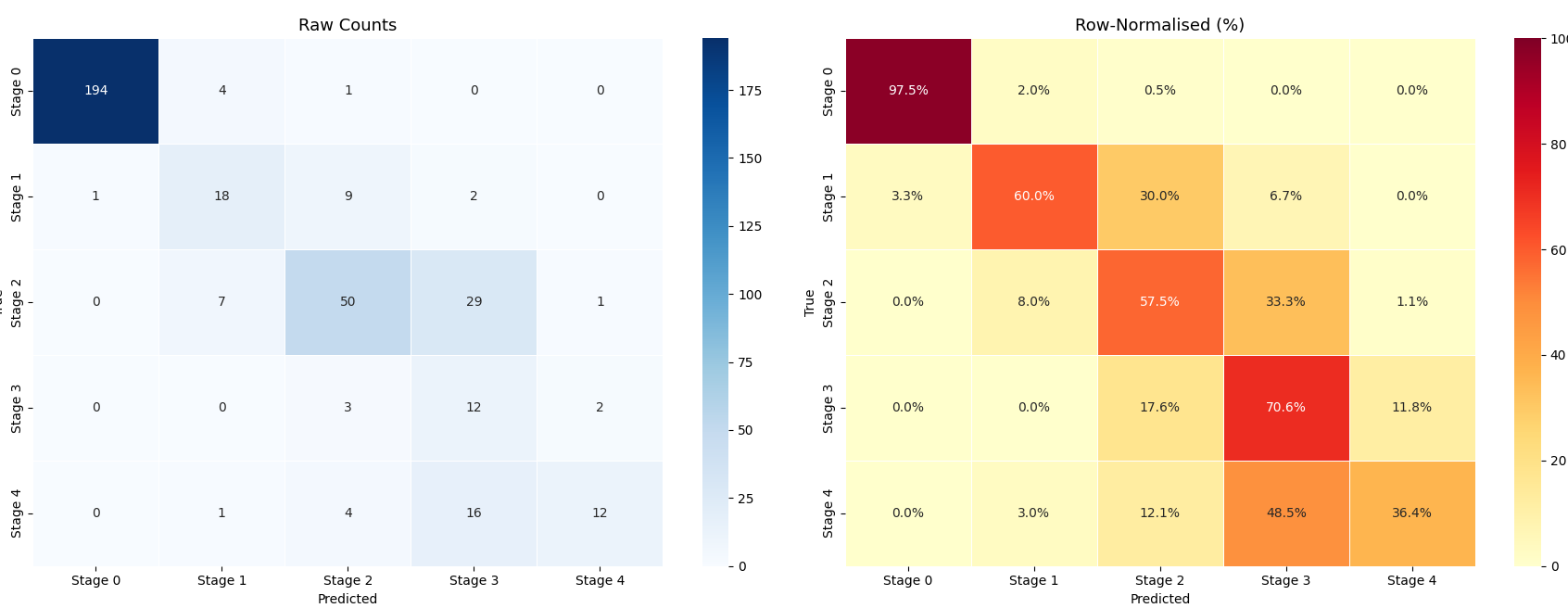}
  \caption{Exploratory confusion matrix on the 366-image analysis
  subset using analysis-selected thresholds. The associated QWK of
  0.9131 is not a held-out estimate.}
  \label{fig:CM}
\end{figure}

\begin{figure}[!htbp]
  \centering
  \includegraphics[width=0.9\linewidth]{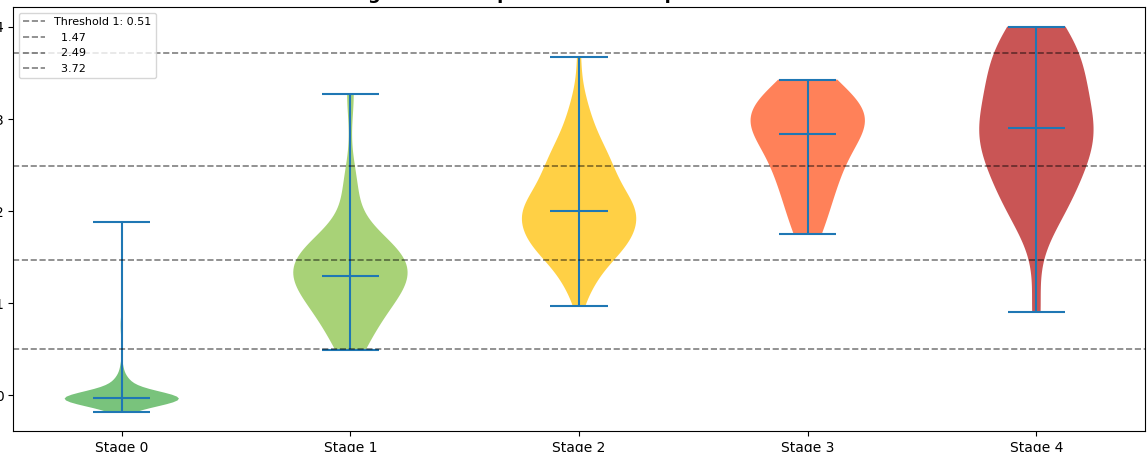}
  \caption{Exploratory raw regression-output distributions by true
  grade. The dashed thresholds were selected using the same analysis-
  subset labels and must not be interpreted as validation-fixed
  decision boundaries.}
  \label{fig:regOutputDistribution}
\end{figure}

The raw-output distributions and the four post-hoc thresholds are
shown in Fig.~\ref{fig:regOutputDistribution}. No confidence interval
is attached to these values because the threshold-selection leakage
invalidates a confirmatory interpretation. A clean rerun should report
fixed-threshold QWK, accuracy, per-grade precision and recall, the
confusion matrix, and bootstrap confidence intervals.

\subsection{Uncertainty Analysis}
\label{subsec:uncertainty}

Mean MC-dropout score variance increases monotonically with DR
severity: mean variance for No-DR is 0.00581, rising through Mild
(0.00844), Moderate (0.01288), Severe (0.02253), to PDR (0.02575).
This association does not demonstrate calibration: variance may also
reflect class frequency, raw-score scale, or the error structure of
this subset. No uncertainty calibration, error-detection AUROC/AUPRC,
area under the risk--coverage curve, distribution-shift test, repeated
seed analysis, or MC-pass convergence analysis is available. The
quantity is therefore used only as a ranking score in the exploratory
selective-prediction analysis.

\subsection{Referral Trade-off}
\label{subsec:referral}

The displayed operating point was selected post hoc by using
analysis-subset score-variance percentiles and QWK, rather than by
locking a numeric $\tau$ on validation data. At the nominal 20\%
referral point, 293/366 images (80.1\%) were accepted and 73/366
(19.9\%) were referred. Although 26/33 PDR images (78.8\%) and 13/17
severe-NPDR images (76.5\%) were referred, the accepted subset still
contained seven PDR and four severe-NPDR images. All seven accepted PDR
cases were missed, and only one of the four accepted severe-NPDR cases
was correct. Thus ten high-grade false negatives remained auto-graded.
The accepted-subset QWK was 0.9040, but it is not directly comparable
with the full-subset QWK because selective removal changes the grade
marginals.

\begin{figure}[!htbp]
  \centering
  \includegraphics[width=0.85\linewidth]{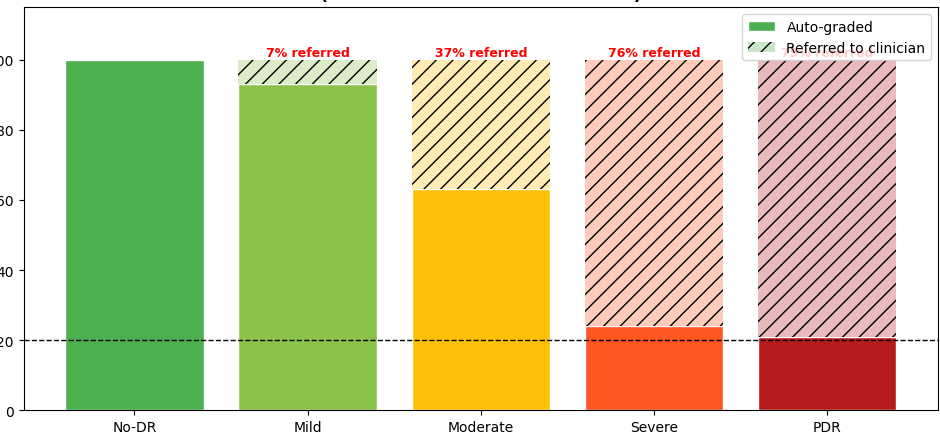}
  \caption{Exploratory grade composition at the post-hoc nominal 20\%
  referral point. The threshold was derived from the same analysis
  subset and is not a locked deployment threshold.}
  \label{fig:uncertainty_by_grade}
\end{figure}

These observations do not establish a safety benefit. A confirmatory
selective-prediction report should give, at every validation-fixed
operating point, coverage, overall error risk, accepted-case accuracy,
accepted-case per-grade sensitivity, fractions of severe/PDR cases
referred, remaining high-grade false-negative counts, and AURC.

\subsection{Grad-CAM Visual Explanations}
\label{subsec:gradcam_results}

\begin{figure}[!htbp]
  \centering
  \includegraphics[width=\linewidth]{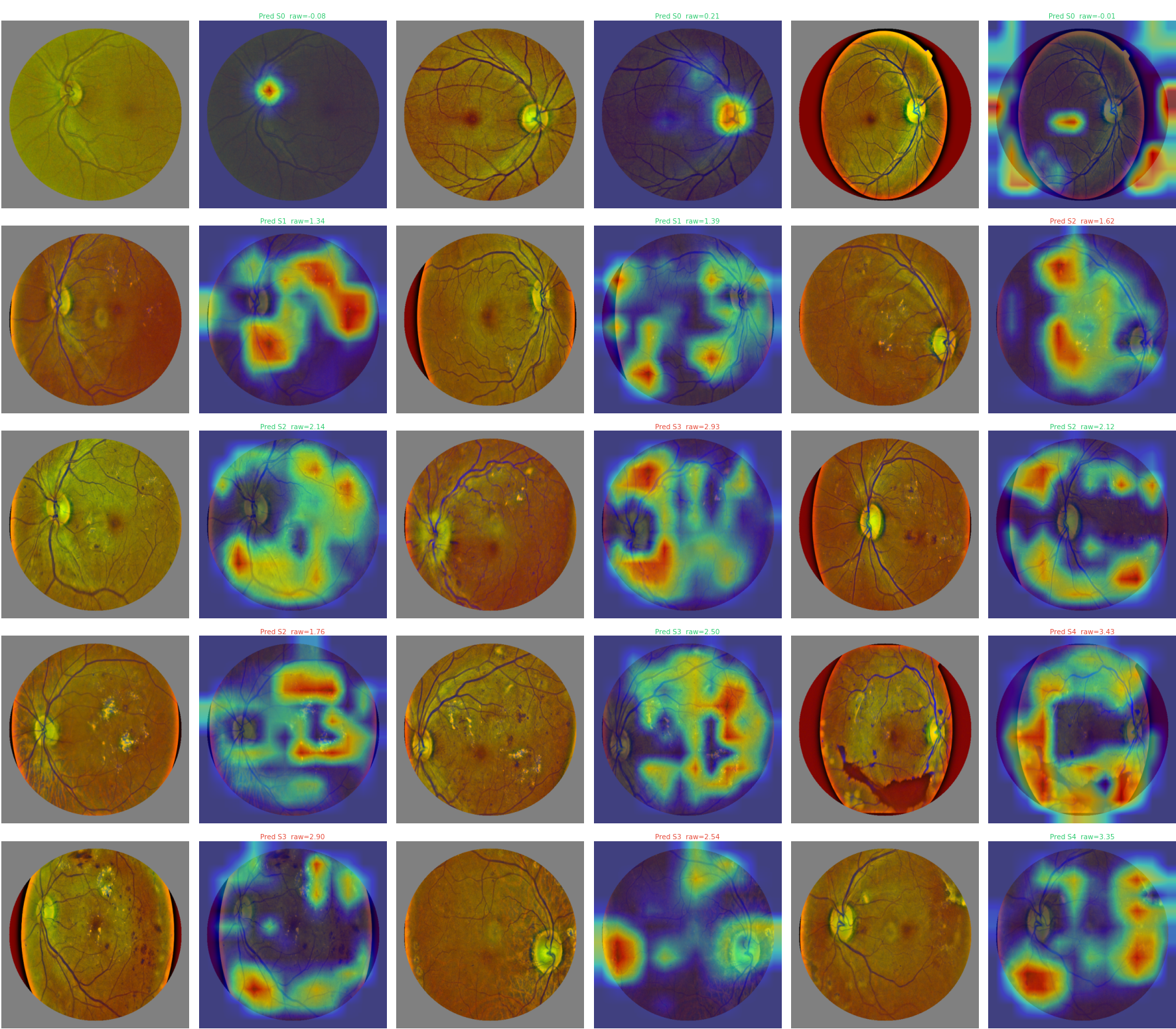}
  \caption{Qualitative Grad-CAM maps. Within each pair, the left image
  is the saved preprocessed feature image and the right image is the
  overlay. Green labels denote a correct exploratory grade and red
  labels an incorrect grade. These maps are not lesion-level
  validation.}
  \label{fig:gradcam}
\end{figure}

Grad-CAM heatmaps computed on the last convolutional block of
EfficientNetV2-L are shown in Fig.~\ref{fig:gradcam}. The examples are
qualitative and were not evaluated against lesion annotations or by
retinal specialists. Several maps are diffuse or emphasise peripheral
regions. Consequently, they neither identify named lesions nor rule out
reliance on imaging artefacts. The sampling rule should be seeded and
reported, and future validation should include failure cases,
lesion-localisation metrics, masked-region perturbation, and
faithfulness tests.

\subsection{Interpretation and Confirmatory Protocol}
\label{subsec:discussion}

Because the current QWK and referral point were selected with
analysis-subset information, they must not be ranked against the
literature values in Table~\ref{tab:qwk_litreview}. The confirmatory
sequence is: (i) correct and ablate the morphology channel; (ii) select
the checkpoint, ordered grade thresholds, and numeric referral
threshold using training/validation data only; (iii) evaluate the
held-out subset exactly once; (iv) report bootstrap confidence
intervals, multiple training seeds, and MC-pass convergence; and (v)
perform external validation, uncertainty calibration/error-detection
analysis, and lesion/explanation validation. Only results from that
locked protocol can support comparative, clinical-safety, or deployment
claims.

\section{Conclusion}
\label{sec:conclusion}

This paper describes a five-stage ordinal DR pipeline and audits the
evidence currently available for it. The audit identifies three issues
that prevent confirmatory interpretation: analysis-subset labels were
used for grade-threshold fitting, the referral operating point was
selected on the same subset, and the stated dark/bright morphological
fusion collapses mathematically to a scaled black-hat response. The
quoted QWK and referral results are therefore retained only as
exploratory diagnostics.

The present evidence also does not establish calibrated uncertainty,
safe accepted-case performance, lesion-level Grad-CAM validity,
cross-dataset generalisation, or deployment readiness. The next step is
a fully reproducible rerun with corrected preprocessing, archived code
and seeds, validation-only model selection, ordered thresholds, a
validation-fixed referral rule, one locked held-out evaluation,
confidence intervals, and external clinical validation.

\section*{Declaration of Competing Interest}
The author declares no competing financial or non-financial interests.

\subsection*{Declaration of generative AI and AI-assisted technologies in the manuscript preparation process.}
During the preparation of this work the author used Anthopic's Claude,OpenAI's ChatGPT and Google's Gemini in order to proofread the language of the manuscript. After using this tool/service, the author reviewed and edited the content as needed and takes full responsibility for the content of the published article.



\bibliographystyle{elsarticle-num-names}
\bibliography{references,extras}

\end{document}